\documentclass[10pt,aps,prd,twocolumn,preprintnumbers]{revtex4-1}

\usepackage{verbatim}
\usepackage{amsmath}
\usepackage{amsfonts}
\usepackage{amssymb}
\usepackage{bbm}
\usepackage{latexsym}
\usepackage{hhline}
\usepackage{color}
\usepackage{graphicx}
\usepackage{hyperref}

\DeclareMathSymbol{\mg}{\mathrel}{symbols}{"1D}

\newcommand{\ga}{\alpha}

\newcommand{\gd}{\delta}

\newcommand{\gvf}{\varphi}

\newcommand{\gk}{\kappa}

\newcommand{\gp}{\pi}

\newcommand{\gL}{\Lambda}

\newcommand{\cV}{{\cal V}}

\newcommand{\ua}{{\underline a}}

\newcommand{\ra}{\rightarrow}

\newcommand{\beq}{\begin{equation}}
\newcommand{\eeq}{\end{equation}}
\newcommand{\barr}{\begin{array}}
\newcommand{\earr}{\end{array}}
\newcommand{\equ}[1]{\begin{gather} #1 \end{gather}}
\newcommand{\equa}[1]{\begin{align} #1 \end{align}}

\newcommand{\arry}[2]{\begin{array}{#1} #2 \end{array}}

\newcommand{\pmtrx}[1]{\begin{pmatrix} #1 \end{pmatrix}}

\newcommand{\non}{\nonumber}
\newcommand{\sfrac}[2]{\mbox{$\frac{#1}{#2}$}}
\newcounter{oldcounter}

\newcommand{\bz}{{\bar z}}
\newcommand{\Intr}{\mathbb{Z}}

\newcommand{\ba}[2]{\[\begin{array}{#2}\label{#1}}
\newcommand{\ea}{\end{array}\]}
\newcommand{\be}{\begin{equation}}
\newcommand{\ee}{\end{equation}}
\newcommand{\bea}{\begin{eqnarray}}
\newcommand{\eea}{\end{eqnarray}}

\newcommand{\sm}{{\,\mbox{-}}}

\begin{document}

%
%
\title{Infinite number of MSSMs from heterotic line bundles? }

\author{Stefan \surname{Groot Nibbelink$^{1}$}}
\email[]{Groot.Nibbelink@physik.uni-muenchen.de}
\author{Orestis \surname{Loukas$^{1,2}$}}
\email[]{O.Loukas@physik.uni-muenchen.de}
\author{Fabian \surname{Ruehle$^3$}}
\email[]{fabian.ruehle@desy.de}
\author{Patrick K.S. \surname{Vaudrevange$^{1,4,5}$,}}
\email[]{Patrick.Vaudrevange@physik.uni-muenchen.de}
\affiliation{$^1$Arnold Sommerfeld Center for Theoretical Physics, Ludwig-Maximilians-Universit\"at M\"unchen, Theresienstra\ss e 37, 80333 M\"unchen, Germany}
\affiliation{$^2$School of Electrical and Computer Engineering, National Technical University of Athens, Zografou Campus, GR-15780 Athens, Greece}
\affiliation{$^3$Deutsches Elektronen-Synchrotron DESY, Notkestra\ss e 85, 22607 Hamburg, Germany}
\affiliation{$^4$TUM Institute for Advanced Study, Lichtenbergstr. 2a, 85748 Garching, Germany}
\affiliation{$^5$Excellence Cluster Universe, Technische Universit\"at M\"unchen, Boltzmannstr. 2, D-85748, Garching, Germany}
\begin{abstract}
We consider heterotic E$_8\times$E$_8$ supergravity compactified on smooth Calabi-Yau (CY) manifolds with line bundle gauge backgrounds. Infinite sets of models that satisfy the Bianchi identities and flux quantization conditions can be constructed by letting their background flux quanta grow without bound. Even though we do not have a general proof, we find that all examples are at the boundary of the theory's validity: the Donaldson-Uhlenbeck-Yau (DUY) equations, which can be thought of as vanishing D-term conditions, cannot be satisfied inside the K\"ahler cone unless a growing number of scalar Vacuum Expectation Values (VEVs) is switched on. As they are charged under various line bundles simultaneously, the gauge background gets deformed by these VEVs to a non-Abelian bundle. In general, our physical expectation is that such infinite sets of models should be impossible, since they never seem to occur in exact CFT constructions.  
\end{abstract}
%
\preprint{DESY-15-082, FLAVOUR(267104)-ERC-102, LMU-ASC 37/15} 
%
%
\maketitle

\section{Introduction}
\label{sc:Introduction}

%
%
String theory is believed to be the ultimate theory of nature unifying all interactions including gravity. The standard string lore says that the number of distinct vacua in the string landscape is extremely huge but nevertheless finite. Estimates range from $10^{500}$ to $10^{1000}$ based on counting D-brane configurations~\cite{Douglas:2003um} and covariant heterotic lattice constructions~\cite{Lerche:1986cx, Schellekens:2014}.

%
%
In light of this, it is very surprising that we observe that it is possible to construct an infinite number of models (that can even be quite similar to the Minimal Supersymmetric Standard Model (MSSM)), which satisfy the fundamental consistency conditions, i.e.\ the appropriate flux quantizations and Bianchi Identities (BIs). 
We explicitly construct infinite sets of line bundle models on smooth Calabi-Yaus in heterotic E$_8\times$E$_8$ supergravity without any NS5-branes or anti-NS5-branes. These models have some unconstrained discrete input parameter, such that the number of states in the hidden sector (i.e.\ the second E$_8$ factor) grows without bound, while the observable sector (i.e.\ the first E$_8$ factor) remains unaltered.
Note that in ref.~\cite{Bouchard:2008bg} infinite sets of non-Abelian bundle models were constructed, however, in the presence of both heterotic NS5-branes and anti-branes. 

Models with arbitrarily large gauge fluxes and growing number of states seem to be unphysical. Hence, we analyze the question at which stage the Effective Field Theory (EFT) description breaks down. However, surprisingly, we do not find a clear bound on the discrete input parameters and we can only judge the EFT validity on a model-by-model basis. 
In the light of this, we also  comment on the results of \cite{Buchbinder:2013dna,Anderson:2010ty}.
Therefore, the allowed parameter range of the discrete input parameters is not sharply determined, which has important consequences for line bundle model-searches.

%
%
This work is structured as follows: We first collect the necessary prerequisites about smooth CY spaces and the construction of line bundles on them. Next, we focus on two particular CY examples to demonstrate how to construct an infinite number of Grand Unified Theory (GUT)-like models that satisfy the fundamental topological consistency conditions. These CYs permit a discrete symmetry that can be used to reduce the GUTs to an infinite set of MSSM-like models with three generations. After constructing these GUT models we critically analyze their EFT validity. In the final section we put the possible implications of our findings for string model building into perspective. 

\section{Smooth Calabi-Yaus}
\label{sc:SmoothCY}

%
%
A CY threefold $X$ is defined as a K\"ahler manifold of three complex dimensions with vanishing first Chern class $c_1=0$. The number of independent complex structure deformations of $X$ is denoted by the Hodge number $h_{21}$. The other Hodge number $h_{11}$ counts the number of linearly independent divisors $D_i$ or equivalently their dual  curves $C_i$, being (complex) codimension one and two subspaces of $X$, respectively. We can choose a basis such that 
\equ{ \label{IntegralBasis} 
\int_{C_i} D_j = \int_{D_i} C_j = \delta_{ij}~, 
}
where we integrate over the Poincar\'e-dual two- and four-forms, respectively. The CY $X$ is further characterized by its triple intersection numbers $\kappa_{ijk}$ and the second Chern class $c_2$ evaluated on the $D_i$\,, i.e.\
\equ{ \label{IntersectionChern}
\kappa_{ijk} = \int_X D_i D_j D_k~,
\quad 
c_{2i}  = \int_{D_i} c_2~.
}
We often use the shorthand notation $\kappa_{ijk} \equiv D_i D_j D_k$ for the triple intersection numbers and write $c_2(D_i) = c_{2 i}$. Volumes of curves $C$, divisors $D$ and the CY $X$ itself are determined by the K\"ahler form $J = a_i\, D_i$\,, 
\equ{
\hspace{-1ex}
\text{Vol}(C) = \!\int_C \!\!J\,,~
\text{Vol}(D) = \sfrac 12\! \int_D \!\! J^2,~
\text{Vol}(X) = \sfrac 16\! \int_ X \!\!J^3,
}
where $a_i$ denote the $h_{11}$ K\"ahler moduli. Here and in the following, summation over repeated indices is implied. Requiring positivity of these volumes defines the K\"ahler/Mori cone. In addition, the supergravity (SUGRA) approximation of string theory can only be trusted if the volumes of all curves, divisors and of the manifold $X$ are sufficiently much larger (denoted by $\gtrsim$) than appropriate powers of the string length $\ell_s$. These two requirements are satisfied if all moduli $a_i$ lie well inside the K\"ahler cone.

\subsection{Complete intersection CY example}

%
%
Complete Intersection Calabi-Yaus (CICYs) are prominent examples of CYs. They are defined as hypersurfaces in products of projective spaces, characterized by polynomials of appropriate scaling degrees such that the CY condition $c_1=0$ is fulfilled. For concreteness we focus on one specific favorable CICY called the tetra-quadric (CICY 7862 in the list given in~\cite{Candelas:1987kf,Braun:2010vc,CYweb}). This manifold has Hodge numbers $h_{11}=4$ and $h_{21}=68$. Furthermore, the non-vanishing triple intersection numbers and the second Chern classes in the basis \eqref{IntegralBasis} are 
\equ{
\gk_{ijk}  = 2
\quad\text{and}\quad 
c_{2 i} = 24~,
}
for $i \neq j \neq k \neq i$ from 1 to 4. For this CICY the SUGRA approximation is valid inside the K\"ahler cone if $a_i \gtrsim \ell_s^2$\,.

\subsection{The Schoen manifold}

%
%
As a second example we discuss the Schoen manifold~\cite{Schoen:1988}. 
Its Hodge numbers are equal, $h_{11} = h_{21} = 19$, and thus the Euler number vanishes. This CY geometry can be described as the blow-up of a particular toroidal $\mathbbm{Z}_2 \times \mathbbm{Z}_2$ orbifold~\cite{Donagi:2008xy}, a fact that leads to a convenient basis for the $h_{11}=3+8+8=19$ independent divisors. Note however that this divisor basis is not minimal in the sense of \eqref{IntegralBasis}. The inherited divisors $R_i$,  $i=1,2,3$, correspond to the 4D sub-tori of the internal six-torus. The $8+8$ exceptional divisors $E_r$ and $\widetilde E_r$ (where $r=r_1r_2r_3$\, is a multi-index with $r_i=0,1$), originate from the blow-up of the $8+8$ orbifold singularities~\cite{Nibbelink:2012de}. The non-vanishing triple intersection numbers and second Chern classes are given by
\begin{subequations}
\equ{
R_1 R_2 R_3 = 4~,~ R_1 E_r E_r = R_2 \widetilde E_r \widetilde E_r = -4~, 
\\[1ex] 
c_2(R_1) = c_2(R_2) = 24~. 
}
\end{subequations} 
For the Schoen manifold the conditions to be well inside the K\"ahler cone are more involved compared to the CICY case discussed above. Roughly, they can be characterized as $\text{Vol}(R_i) > 4 \text{Vol}(\widetilde{E}_r) \gtrsim 4\, \ell_s^4$ and similarly for the divisors $E_r$. 

\section{Infinite sets of models allowed by the Bianchi identities}
\label{sc:InfinteModels}
In this section we use the two CY examples to construct line bundle backgrounds where the Bianchi identities and the flux quantization conditions allow in principle for an infinite number of line bundle models.

%
%
In general, on some CY $X$ we can construct a gauge background $\mathcal{F}$ that is parametrized by a sum of line bundles
\equ{  \label{LineBundleFlux} 
 \frac{\mathcal{F}}{2\pi} = D_i\, H_i~, 
 \quad 
 H_i = V_i^I\, H_I~. 
 }
Each constant 16-component line bundle vector $V_i$ characterizes how the line bundle, supported on the divisor $D_i$, is embedded in the Cartan subalgebra of the 10D gauge group E$_8\times$E$_8$. $H_I$ with $I=1,\ldots,16$ denotes the 16 Cartan generators and $\gL$ the root lattice of $\text{E}_8\times\text{E}_8$. On any string state, characterized by left-moving momenta $p \in \gL$, $H_I$ is evaluated as $H_I(p) = p_I$, hence $H_i(p) = V_i \cdot p$. We also use the notation $V_i = (V^{\prime}_i, V^{\prime\prime}_i)$ for the decomposition of bundle vectors $V_i$ into observable and hidden E$_8$ factors respectively, and similarly for any other relevant object. 

%
%
The flux background~\eqref{LineBundleFlux} has to be integrally quantized on all string states when integrated over any curve $C_i$ in the basis \eqref{IntegralBasis}. Using \eqref{LineBundleFlux} this implies that all bundle vectors have to lie on the $\text{E}_8\times\text{E}_8$ root lattice, i.e.\ 
\equ{  \label{LineBundleFluxQuantization} 
V_i \in \Lambda~. 
}
%

%
%
The integrated BIs of the Kalb-Ramond field in the presence of NS5 branes read
\begin{subequations}
 \label{BIs}
\equ{
N_i = N^{\prime}_i + N^{\prime\prime}_i\,, \\[1ex] 
\label{NS5charges} 
N^{\prime}_i =  c_{2i} + \kappa_{ijk}\, V^{\prime}_j \cdot V^{\prime}_k~,~N^{\prime\prime}_i =  c_{2i} + \kappa_{ijk}\, V^{\prime\prime}_j \cdot V^{\prime\prime}_k
}
\end{subequations}
on all divisors $D_i$\,, $i=1,\ldots, h_{11}$. The BIs are vital for anomaly cancellation in the 4D EFT. Unbroken SUSY requires that the NS5-brane charges $N_i$ are non-negative.
We focus on models without NS5-branes, in which case \eqref{BIs} reads $N_i = 0$, so that $N^{\prime\prime}_i= -N^{\prime}_i$.

%
%
Next, we review how the 4D gauge group and the (chiral) matter spectrum can be computed. In 10D the E$_8\times$E$_8$ gauge group is specified by a set of $240+240$ roots \{$p\in \gL$ with $p^2=2$\}. Then, having defined the gauge background~\eqref{LineBundleFlux}, the 4D gauge group is given by those roots $p$ that are uncharged under $\mathcal{F}$, i.e.\ 
\equ{ \label{GaugeGroup}
H_i(p) = V_i \cdot p ~=~ 0 \quad\forall~ i = 1,\ldots,h_{11}~.
}
Furthermore, the chiral part of the 4D matter spectrum can be computed using the multiplicity operator, 
\equ{ \label{MultiOp}
\mathcal{N}  = \sfrac16\, \kappa_{ijk}\, H_i H_j H_k + \sfrac1{12} c_{2i}\,  H_i~, 
}
evaluated on each root $p$. A left-chiral 4D matter state with weight $p$ has positive multiplicity $\mathcal{N}(p) \in \mathbbm{N}$, while a right-chiral CPT partner has negative multiplicity, since $\mathcal{N}(-p) = -\mathcal{N}(p)$.

%
%
In refs.~\cite{Anderson:2011ns,Anderson:2012yf,Anderson:2013xka} a different notation for the line bundle embedding into a single E$_8$ is used: A vector bundle $\mathcal{V}$ with structure group S(U(1)${}^5$) is constructed on the CY $X$ as a direct sum of line bundles
\equ{ 
\mathcal{V} = \bigoplus_{a=1}^5
 \mathcal{O}_X\Big(k_{(a)}^1,\ldots, k_{(a)}^{h_{11}}\Big)~, 
\quad 
\sum_{a=1}^5 k_{(a)}^i = 0~, 
}
labeled by the vectors $k_{(a)}= (k_{(a)}^1, \ldots, k_{(a)}^{h_{11}})\in\Intr^{h_{11}}$ with $a=1,\ldots,5$ and $i=1,\ldots,h_{11}$. The corresponding bundle vectors $V_i^\prime = (a_i^5, b_i, c_i, d_i)$ (exponents indicate repetition of these entries) in the observable E$_8$ are given by 
\equ{  \label{SU5inE8emb} 
\pmtrx{ a_i \\ b_i \\ c_i \\ d_i} = 
- \sfrac 12 
\pmtrx{ 
~~1 & ~~1 & ~~1 & ~~1 \\ 
~~1 & -1 & -1 & ~~1 \\ 
-1 & ~~1 & -1 & ~~1 \\
-1 & -1 & ~~1 & ~~1 
}
\pmtrx{ k_{(1)}^i \\ k_{(2)}^i \\ k_{(3)}^i \\ k_{(4)}^i}~,
}
for $i=1,\ldots,h_{11}$ and using $k_{(5)} = -k_{(1)}-k_{(2)}-k_{(3)}-k_{(4)}$.

%
%
\begin{table}
\begin{center}
\renewcommand{\arraystretch}{1.2}
\begin{tabular}{|c|c||c|c|}
\hline
\multicolumn{2}{|c||}{Observable E$_8$} & \multicolumn{2}{c|}{Hidden E$_8$}\\
~~Mult.~~ & ~~~Rep.~~~  &  ~~Mult.~~ & ~~~Rep.~~~
\\\hline\hline 
12 & $(\overline{\textbf{5}},\textbf{1})$ & 12k + 8 & $(\textbf{1},\textbf{4})$\\
12 & $(\textbf{10},\textbf{1})$           & 12k + 8 & $(\textbf{1},\overline{\textbf{4}})$\\
   &                                      &      4k & $(\textbf{1},\textbf{6})$\\ 
 60  & $(\textbf{1},\textbf{1})$            & 80k + 8 & $(\textbf{1},\textbf{1})$\\ \hline
\end{tabular}
\renewcommand{\arraystretch}{1}
\end{center}
\caption{\label{tb:CICY7862Spectrum}
SU(5) GUT models with 12 generations from the CICY 7862: The charged spectrum, obtained from \eqref{CICY7862BundleVectors}, is summarized for $k \in \mathbbm{N}$ with gauge group $\text{SU}(5)\times\text{SU}(4)$ (omitting the U(1)$^{9}$ charges). The multiplicities in the observable sector are independent of $k$, while they grow linearly with $k$ in the hidden sector. 
}
\end{table}

\subsection{Infinite models for the CICY example}

%
%
For the CICY 7862 the line bundle vectors $V_i$ have to be integrally quantized in the sense of \eqref{LineBundleFluxQuantization} and the four BI identities~\eqref{BIs} without NS5-branes read 
\equ{ \label{BIsCICY} 
24\, + \sum_{(i \neq j) \neq k} V_i \cdot V_j = 0~.
}
for each $k=1,\ldots,4$. 

%
%
Now, we explicitly construct an example of an infinite set of CY models: We start from a specific 12 generation SU(5) GUT model constructed in~\cite{Anderson:2011ns,Anderson:2011_data} with:
\equ{ \label{CICY7862ChargesK}
\Big( k^i_{\ (a)} \Big) =  
\left(\arry{rrrrr}{
1 & 1 & 0 &- 1 &  -1 \\[-.5ex]
-1 & -2 & 0 & 2 & 1 \\[-.5ex]
-1 & 0 & -1 & 2 & 0 \\[-.5ex]
1 & 1 & 1 & -3 & 0
}\right)~. 
}
An order four symmetry can be used to reduce this to an MSSM-like model with three generations. Using the prescription~\eqref{SU5inE8emb}, the four bundle vectors $V_i$ are given by
\equ{ \label{CICY7862BundleVectors}
\arry{lcl}{
V_1 &=&
( \sm\sfrac12^5, ~\,\sfrac12, ~\,\sfrac12, ~\,\sfrac32\,)( ~\, 1^4,  ~\,0,    \sm2, ~\, 0, ~\,0 \,)~, 
\\[.5ex] V_2 &=&
( ~\,\sfrac12^5,  \sm\sfrac32,  \sm\sfrac12,  \sm\sfrac52\,)(~\, 0^4,    \sm1, ~\,1,  ~\, 0,  2k \,)~, 
\\[.5ex]  V_3 &=&
( ~\,0^5, \,\sm1,  \,\sm2, \,\sm1\,)(  \sm1^4,  ~\,1,  ~\,1, ~\, 0, ~\,0 \,)~, 
\\[.5ex] V_4 &=& 
( ~\,0^5,  ~~2, ~~2, ~~2\,)(~\, 0^4,     ~\,0,   ~\,0, ~\, 0,  ~\,0 \,)~.
}
}
Here, we have amended the original $V_i^{\prime}$ of \eqref{CICY7862ChargesK} by vectors $V_i^{\prime\prime}$ in the hidden sector (parametrized by $k \in \mathbbm{N}$) such that the NS5-brane charges~\eqref{BIs} vanish for all divisors: 
\equ{ \label{CICY7862NS5charges}
\begin{array}{lcl}
N^{\prime}_1 = -N^{\prime\prime}_1 =  -24~, 
& \quad & 
N^{\prime}_2 = -N^{\prime\prime}_2 = 0~, 
\\[0.5ex]
N^{\prime}_3  = -N^{\prime\prime}_3 = -16~, 
& \quad  &
N^{\prime}_4 = -N^{\prime\prime}_4  = 8~,
\end{array}
}
and the BIs~\eqref{BIsCICY} are fulfilled. Using the multiplicity operator \eqref{MultiOp}, the resulting chiral spectra are listed in Table~\ref{tb:CICY7862Spectrum}. Since $V_i^{\prime}$ is taken from~\eqref{CICY7862ChargesK}, the observable spectrum agrees with the one of \cite{Anderson:2011ns,Anderson:2011_data}. We have also computed the non-chiral spectra by investigating the dimensions of the corresponding sheaf cohomology groups using \texttt{cohomcalg} \cite{Blumenhagen:2010pv,cohomCalg:Implementation}. In addition to the vector-like pairs inherited from the visible sector of \cite{Anderson:2011_data}, we find new vector-like pairs in the hidden sector. In all cases (except for one in which the dimensions are not uniquely fixed by the Koszul sequence) their multiplicities are independent of $k$.

%
%
Using this example, we are now in the position to explain why it is possible to generate an infinite number of line bundle models. The line bundle vectors~\eqref{CICY7862BundleVectors} include an arbitrary integer $k>1$. However, the NS5-brane charges~\eqref{CICY7862NS5charges} do not depend on $k$: Given the intersection numbers of the CICY~7862, $N^{\prime}_i$ and $N^{\prime\prime}_i$ only involve inner products $V^\prime_i \cdot V^\prime_j$ and $V^{\prime\prime}_i \cdot V^{\prime\prime}_j$ for $i\neq j$ in which the $k$-dependence drops out. However, the multiplicity operator~\eqref{MultiOp} depends on $k$ as can be seen in Table~\ref{tb:CICY7862Spectrum}. Furthermore, we have confirmed that all pure and mixed (non-) Abelian gauge and gravitational anomalies cancel independently of $k$ by taking the generalized Green-Schwarz mechanism~\cite{Blumenhagen:2005ga} into account.

%
%
Even though we have presented only a very specific example of an infinite set of models, our findings can be extended to other CICYs as well, since the effect only relies on the fact that the BIs \eqref{BIs} can be insensitive to some of the line bundle data. In particular, the norm of some (in the presented example all) line bundle vectors is left fully unconstrained by the integrated BIs. Hence, whenever one allows breaking the hidden E$_8$, infinite sets of MSSM-like models seem to be possible. If one restricts oneself only to MSSMs with unbroken hidden E$_8$, the number of GUT representations will generically depend on the full line bundle data.

%
%
\subsection{Infinite models for the Schoen manifold}

In order to illustrate the effect for another class of CYs we now construct an infinite set of models on the Schoen manifold. 
Denoting the 19 line bundle vectors supported on the divisors  $R_i, E_r, \widetilde E_r$ by $B_i, V_r$ and $\widetilde V_r$\,, the BIs~\eqref{BIs} without NS5-branes read: 
\equ{ \label{BIsSchoen} 
B_1 \cdot V_r = 0~, 
\quad 
B_2 \cdot \widetilde V_r  = 0~, 
\quad 
B_1 \cdot B_2 = 0~, 
\\[1ex]  \non 
\sum_r (V_r)^2 = 12 + 2\, B_2\cdot B_3~,~
\sum_r (\widetilde V_r)^2 = 12 + 2\, B_1\cdot B_3~,
}
subject to the flux quantization conditions given in \cite{Nibbelink:2012de}. For this geometry, the BIs on the second line allow us to construct an infinite set of solutions by increasing, for example, the lengths of the vectors $\widetilde V_r$ and the inner product $B_1\cdot B_3$ simultaneously such that their contributions cancel.

%
%
To be concrete we consider the flux background $\mathcal{F}$ specified by the bundle vectors 
\equ { \label{Schoen_BundleVectors}
\begin{array}{lrl}
B_1 &=& (\sm\sfrac 12^2, \sfrac 12, \sm\sfrac 12^4,  \sfrac 12)(0^3, 1, \sm 1,  0,  2k,  2k)~, 
\\[1ex] 
B_2 &=& (0^8)(0^6,  \sm2k,  2k)\,, 
\\[1ex] 
B_3 &=& (\sfrac 12^2, \sm\sfrac 12, \sfrac 12^4, \sm\sfrac 12)(0^3,  \sm 1,  1,  0^2,  2k)~, 
\\[1ex] 
V_{0r_2r_3} &=& p (\sm\sfrac 12^3, \sfrac 12,  \sm 1, 0^3)(\sm\sfrac 12^4, 0,  (p+1)k,  0^2)~, 
\\[1ex] 
\widetilde V_{0r_2r_3} &=& p (~\,\sfrac 12, \sm\sfrac 12, 0^6)(0^3,  \sfrac 12^2,  \sm (p+1)k,  0^2)~,
\end{array} 
}
where $p = (-1)^{r_2+r_3}$ and $V_{1r_2r_3} = \widetilde V_{1r_2r_3} = 0$. This class of models has gauge group $\text{SU}(5)\times\text{SU}(3)\times\text{U}(1)^{10}$. The charged matter spectrum is given in Table~\ref{tb:SchoenSpectrum}. Again, all anomalies cancel independently of the parameter $k \in \mathbbm{N}$ which enumerates the infinite set of models.

%
%
If we set the magnetic fluxes $B_i$ on the tori to zero, the number of models is finite. Indeed, the non-trivial BIs then read:\ $\sum_r (V_r)^2 = \sum_r (\widetilde V_r)^2 = 12$. These conditions constrain the lengths of the bundle vectors, $(V_r)^2 \leq 12$, and hence only admit a finite set of solutions. 
This case can be interpreted as a blow-up of a $\mathbbm{Z}_2 \times \mathbbm{Z}_2$ orbifold without magnetic flux on the six-torus~\cite{Nibbelink:2012de}: The $8+8$ massless blow-up modes (without string oscillator excitations), resolving the $8+8$ orbifold singularities, carry massless left-moving momenta, characterized by the $8+8$ bundle vectors $V_r$ and $\widetilde{V}_r$ of equal length $(V_r)^2 = (\widetilde V_r)^2 = 3/2$.

%
%
\begin{table}
\begin{center}
\renewcommand{\arraystretch}{1.2}
\begin{tabular}{|c|c||c|c|}
\hline
\multicolumn{2}{|c||}{Observable E$_8$} & \multicolumn{2}{c|}{Hidden E$_8$}\\
~~Mult.~~ & ~~~Rep.~~~  &  ~~Mult.~~ & ~~~Rep.~~~
\\\hline\hline 
6 & $(\textbf{5},\textbf{1})$  & 
$8k(12 k^2+1)+4$ & $(\textbf{1},\overline{\textbf{3}})$
\\
6 & $(\overline{\textbf{10}},\textbf{1})$ & 
$8k(12 k^2+1)+4$ & $(\textbf{1},\textbf{3})$ 	 
\\
66& $(\textbf{1},\textbf{1})$  & 
$8k(40k^2+16k+5)$ & $(\textbf{1},\textbf{1})$        
\\ \hline
\end{tabular}
\renewcommand{\arraystretch}{1}
\end{center}
\caption{\label{tb:SchoenSpectrum} 
SU(5) GUT models with six generations from the Schoen manifold~\eqref{Schoen_BundleVectors}: The charged spectrum is summarized for $k \in \mathbbm{N}$ with gauge group $\text{SU}(5)\times\text{SU}(3)$ (omitting the U(1)$^{10}$ charges). An order two symmetry can be used to reduce this to MSSM-like models with three generations. Again the multiplicities in the observable sector are independent of~$k$. 
}
\end{table}

\section{Breakdown of the EFT description for infinite sets of models}
\label{sc:Breakdown} 

In this section we study the examples of infinite sets of line bundle models, discussed in the previous section, in more detail in order to determine at which point the EFT approach to smooth heterotic models breaks down.

%
%
First of all, one has to ensure that the gauge couplings remain finite such that the theory stays perturbative. 
The reduction of the Green-Schwarz mechanism leads to non-trivial axion couplings which are encoded in the gauge kinetic functions. In the absence of NS5-brane charges the resulting gauge couplings in the observable E$_8$ factor read~\cite{Blumenhagen:2005ga,Blumenhagen:2006ux}: 
\equa{\label{GaugeCouplings}
\frac{2 \pi}{(g^{\prime})^2} &= 
\frac{2}{e^{2\gvf}} 
- \frac{N_i^\prime}{2}\, \frac{a_i}{\ell_s^2}~,
\\[1ex] \non 
\frac{2 \pi}{(g^{\prime})^2_{IJ}} & = 
\frac{2}{e^{2\gvf}}\, \gd^\prime_{IJ}
-  \frac{a_i}{\ell_s^2}
\Big( 
\frac{N_i^\prime}{2}\, \gd^\prime_{IJ} 
+ \frac 23\, \gk_{ijk}\, V_j^{\prime\, I} V_k^{\prime\, J}
\Big)~,
}
for non-Abelian and Abelian gauge group factors, respectively. $\gd^\prime_{IJ}$ denotes the Kronecker delta for the Cartan indices in the first E$_8$. 
We have absorbed $\text{Vol}(X)$ in the definition of the 4D dilaton $\gvf$. 
For the gauge couplings in the hidden E$_8$ one has to interchange ${}^\prime$ and ${}^{\prime\prime}$. In addition, there is kinetic mixing for U(1)s between both E$_8$ factors.

In the examples discussed above we have always ensured that $N_i^\prime =- N_i^{\prime\prime}$ are independent of the scaling parameter $k$ for all $i$\,. Hence, the gauge couplings of non-Abelian gauge group factors remain perturbative when $k$ is increased.
However, given that some components of the bundle vectors $V_i^{\prime\prime}$ in the hidden sector grow with $k$, we see from \eqref{GaugeCouplings} that $(g^{\prime\prime})_{IJ}$ will generically depend on $k$\,. In particular, this can lead to strong coupling for some U(1)s as $k$ becomes large. To avoid such a behavior, we either have to make the corresponding K\"ahler moduli $a_i$ smaller than $\ell_s^2$ or we have to choose $e^{-2\gvf}$ very large. The former option clashes with the validity of the SUGRA approximation while the latter option would turn off the non-Abelian gauge couplings, e.g.\ for SU(5).

%
%
Furthermore, the gauge background has to satisfy the DUY  equations, which can be interpreted as D-flatness conditions for the corresponding U(1)s. This observation has been used to derive the one-loop correction to the DUY equations~\cite{Blumenhagen:2005ga,Blumenhagen:2006ux} and to incorporate the contributions of charged non-Abelian singlet VEVs $\langle s^\prime\rangle$:
\equ{   \label{DUYwithVEVs} 
\hspace{-1ex} 
Q_{ix}^\prime\, \frac{\text{Vol}(D_i)}{\ell_s^6} 
- 
Q_x^\prime\,  \frac{e^{2\gvf}}{8\pi^2}\,  \frac{\text{Vol}(X)}{\ell_s^8}
 + 
\sum q_x^\prime 
 \big| \langle s^\prime \rangle \big|^2 
 = 0~, 
}
using 
$Q_x^\prime = \frac{N_i^\prime}4\, Q_{ix}^\prime$ 
and 
$Q_{ix}^\prime = V^\prime_i \cdot t^\prime_{(x)}$ 
in the absence of NS5-branes, $N_i=0$. Furthermore, the charge $q_x^\prime$ of a state $s^\prime$ (with weight $p^\prime$) is defined w.r.t.\ the $x$-th U(1) generator as $q_x^\prime = t^\prime_{(x)} \cdot p^\prime$ in the observable sector and we have absorbed the field space metric in the definition of $|\langle s^\prime \rangle|^2$. Again, for the hidden sector ${}^\prime$ and ${}^{\prime\prime}$ have to be interchanged.

%
%
Without the singlet VEVs the DUY equations~\eqref{DUYwithVEVs} can be read as a condition on the divisor volumes $\text{Vol}(D_i)$ and thus on the K\"ahler moduli $a_i$\,. Nevertheless, for some choice of bundle vectors $V_i$ the DUY conditions may set some $a_i$ to be close to or at the boundary of the K\"ahler cone. In order to avoid this, the DUY equations do not only affect the moduli but also constrain the bundle vectors $V_i$\,. One may consider balancing the contributions from the divisor volumes with the one-loop correction in \eqref{DUYwithVEVs} while setting $\langle s^\prime \rangle = 0$\,, see e.g.~\cite{Anderson:2010ty,Anderson:2011cza}. However, this typically leads to strong coupling effects, see \eqref{GaugeCouplings}, since in such a case the K\"ahler moduli $a_i$ and the dilaton $\gvf$ have to be related to make the one-loop contribution comparable to the tree-level.

%
%
Alternatively, one can switch on VEVs of charged fields, $\langle s^\prime \rangle \neq 0$. This signifies that the line bundle background has to be deformed to a non-Abelian vector bundle. If such VEVs are necessary to fulfill the DUY conditions inside the K\"ahler cone, the resulting non-Abelian bundle is characterized as non-split~\cite{Anderson:2010ty}. In order that it is reliable to first use the line bundle background to determine the chiral spectrum and then be forced to deform away from it, it is necessary that one can continuously take the limit in which the gauge coupling tends to zero so that the D-term potential vanishes. Moreover, to really guarantee that this configuration corresponds to a bona fide non-Abelian background, one should show that the resulting bundle satisfies the full non-Abelian DUY equations, i.e.\ confirm that it is poly-stable. This analysis would require constructing the non-Abelian bundle corresponding to the deformed Abelian bundle with VEVs. If a description in terms of a monad bundle exists, bundle stability could be checked using the techniques of e.g.~\cite{Anderson:2008uw,Buchbinder:2013dna}. However, this analysis is beyond the scope of the current paper.

%
%
Instead, we focus on some necessary conditions on the VEVs such that the corresponding non-Abelian bundle has any chance at all to be poly-stable. By rewriting the 10D action in terms of 4D N=1 superfields, see e.g.~\cite{Marcus:1983wb,ArkaniHamed:2001tb}, the 4D components $A_\mu$ of the 10D E$_8\times$E$_8$ gauge fields become parts of 4D vector multiplets with auxiliary fields~$D$. Their internal components $A_a$ (using holomorphic coordinates $z^{a}$ and their conjugates $\bz^{\ua}$ on the CY) form chiral superfields; we denote their auxiliary fields by $f_{ab} = -f_{ba}$. Using this formalism one can interpret the Hermitian Yang-Mills equations as vanishing F- and D-terms, i.e.
\equ{ \label{FandDfrom10D}
f_{ab} = F_{ab} =0~, 
\quad 
D = G^{\ua a}\, F_{a\ua} = 0~, 
}
taking values in the adjoint of the full E$_8\times$E$_8$. Here, $G^{\ua a}$ denotes the inverse K\"ahler metric on the CY. The VEVs of the charged scalars are ultimately just internal components of the E$_8\times$E$_8$ gauge fields, and thus contribute to \eqref{FandDfrom10D}. These F-terms can partially be taken into account in the 4D EFT by considering the most general superpotential of these charged scalars. Concerning D-flatness one typically takes only those 4D D-terms into account which correspond to the gauge directions that commute with the (original) bundle background. Hence, from \eqref{FandDfrom10D} we see that this approach misses additional conditions due to D-terms in the other E$_8\times$E$_8$ directions and F-terms in the commuting gauge directions.  Furthermore, even though ref.~\cite{Blumenhagen:2005ga} describes one-loop corrections to the Abelian part of the DUY conditions in 4D, see~\eqref{DUYwithVEVs}, it is not clear to us whether this description includes all one-loop corrections to the 10D equations~\eqref{FandDfrom10D}.

%
%
Finally, we neglect the full backreaction of the internal flux on the geometry. The DUY equations partly account for this backreaction by linking the number of flux quanta to the CY volume, cf.\ \eqref{DUYwithVEVs}. The line bundle backgrounds though, are gauge fluxes that also appear in higher-dimensional operators which can normally be neglected in the $\ga'$-expansion. However, in the limit of arbitrarily large fluxes, the latter become dominant and the EFT description of string theory necessarily falls apart.

%
%
\subsection{EFT Breakdown for the infinite set of CICY 7862 models}

%
%
The original model~\cite{Anderson:2011ns} without the hidden sector bundle lies deep inside the K\"ahler cone, as all the K\"ahler parameters can be taken equal $a_i= a$ (at tree level) and larger than the string scale $a \gtrsim \ell_s^2$. After the hidden bundle extension \eqref{CICY7862BundleVectors} characterized by $k>1$ is switched on, we still want to be inside the K\"ahler cone and in the perturbative regime of the gauge couplings.

%
%
The NS5-brane charges~\eqref{CICY7862NS5charges} do not depend on $k$, as discussed in the previous section. This means that the non-Abelian gauge couplings (first line in \eqref{GaugeCouplings}) can be kept positive and inside the perturbative regime for every $k$. On the other hand, the mixed U(1) gauge couplings (second line in \eqref{GaugeCouplings}) are quickly driven into the non-perturbative regime, unless we let the dilaton depend on $k$ as well,
\equ{ \label{DilScalingCICY7862} 
\frac{1}{e^{2\gvf}}  \gtrsim 8\, k\, \frac{a}{\ell_s^2}~.
}
This estimate is obtained by investigating the smallest eigenvalue of the gauge kinetic function matrix~\eqref{GaugeCouplings} in the large $k$ limit. Phenomenologically, the consequences of \eqref{DilScalingCICY7862} are not very attractive in the large $k$ limit, as \eqref{GaugeCouplings} implies that we tend to ultra-weak coupling, in particular, for the observable SU(5) gauge coupling.

%
%
Next, we discuss the DUY conditions~\eqref{DUYwithVEVs}, i.e.\ the D-terms of U(1)s. First, to accommodate the one-loop correction to the DUYs in the hidden sector, we choose divisor volumes
\equ{ \label{DivVolCICY7862}
\text{Vol}(D_i) = 6\, a^2 
+   \frac 14\,{N_i^{\prime\prime}}\, \frac{e^{2\gvf}}{8\pi^2}\, \frac{\text{Vol}(X)}{\ell_s^2}~.
}
This means in particular that the K\"ahler parameters $a_i$ are not equal, once one-loop effects are taken into account. However, in light of~\eqref{DilScalingCICY7862}, we may assume that all K\"ahler parameters can be taken equal, when $k$ becomes large.

%
%
\begin{table}
\begin{center}
\scalebox{.9}{
\renewcommand{\arraystretch}{1}
\begin{tabular}{|c|c|c||c|c|c|}
\hline
\multicolumn{3}{|c||}{Observable singlets} & 
 $4$ & ~0,~0,-2,-2,~0 & $s^{\prime\prime}_{3}$
\\
~~Mult.~~ & U$^\prime$(1)$^{4}$ charges  & Label & 
 $4$ &  ~0,~0,-2,~2,~0 & $s^{\prime\prime}_{4}$
\\\cline{1-3} \multicolumn{3}{c|}{} &&& \\[-3ex]\cline{1-3}
  $4$  & ~0,~0,~2,-2 & $s^\prime_{1}$ &
 $4k$ &  ~0,-2,~0,~0,~2 & $s^{\prime\prime}_{5}$
 \\
 $12$ & ~0,~0,~2,~2 & $s^\prime_{2}$ &
 $4k$ & ~0,~0,~0,-2,~2 & $s^{\prime\prime}_{6}$
\\
 $12$ & ~5,-1,~1,-1 & $\langle s^\prime_{3} \rangle$ &
 $4k$ &  ~0,~0,~0,~2,~2 & $s^{\prime\prime}_{7}$
\\
 $12$ & ~5,~1,~1,~1 & $\langle s^\prime_{4} \rangle$ & 
 $4k$ & ~0,~2,~0,~0,~2 & $s^{\prime\prime}_{8}$
\\
 $20$ & ~0,~2,~2,~0 & $\langle s^\prime_{5} \rangle$ &
$10k-12$ & -2,-1,~1,~1,-1 & $s^{\prime\prime}_{9}$
\\ 
&&& 
 $10k+12$ & ~2,~1,-1,~1,-1 & $s^{\prime\prime}_{10}$
\\\cline{1-3} \multicolumn{3}{c|}{} &&& \\[-3ex]\cline{1-3}
\multicolumn{3}{|c||}{Hidden singlets} &  
 $16k$ & -2,~1,~1,-1,-1 & $s^{\prime\prime}_{11}$
\\
~~Mult.~~ & U$^{\prime\prime}$(1)$^5$ charges & Label & 
 $2k-4$ & ~2,-1,~1,~1,-1 & $s^{\prime\prime}_{12}$
\\\cline{1-3} \multicolumn{3}{c|}{} &&& \\[-3ex] \cline{1-3}
 $4k-4$ & ~~~0,~0,~2,~0,-2~~ & 	$\langle s^{\prime\prime}_{1} \rangle$ & 
 $16k$ &  ~~~2,-1,-1,-1,-1~~ & 	$s^{\prime\prime}_{13}$ 
\\ 
 $4k+4$ &  ~0,~0,-2,~0,-2 & 	$\langle s^{\prime\prime}_{2} \rangle$ &
 $2k+4$ &   -2,~1,-1,~1,-1 & 	$s^{\prime\prime}_{14}$  
\\ \hline
\end{tabular}
\renewcommand{\arraystretch}{1}
}
\end{center}

\caption{\label{tb:CICY7862Singlets} 
Detailed spectrum of charged non-Abelian singlets with their multiplicities. We distinguish them by whether they are charged under the four observable or five hidden U(1)s and indicate only the relevant charges using the U(1) basis~\eqref{CICYU1Basis}. The fields that take VEVs are denoted by~$\langle s \rangle$. 
}
\end{table}

%
%
Next, we may switch on VEVs for some charged non-Abelian singlets in both the observable and the hidden sectors: The VEVs of the observable singlets take care of the one-loop correction to the DUY equations for the U(1)$^4$ factors from the first E$_8$ while the VEVs of the hidden singlets have to guarantee D-flatness only at tree-level for the U(1)$^5$ factors from the second E$_8$ (since the one-loop correction is trivially satisfied there due to our ansatz~\eqref{DivVolCICY7862}).
All non-Abelian singlets and their observable and hidden charges are listed in Table~\ref{tb:CICY7862Singlets} using the U(1) basis 
\begin{subequations} \label{CICYU1Basis}
\equa{ 
\hspace{-.5ex}
t^\prime_{(0)} &= (2^5, 0^3)~,~
t^\prime_{(x)} = (0^5, 0^{x-1}, 2, 0^{3-x} )~,~
\\[1ex] 
\hspace{-1ex}
t^{\prime\prime}_{(0)} &= (1^4, 0^4)~,~
t^{\prime\prime}_{(y)} = (0^4, 0^{y-1}, 2, 0^{4-y} )~,
}
\end{subequations}
for $x=1,2,3$ and $y=1,\ldots,4$. To be precise, an example of a D-flat and (to all perturbative orders) F-flat VEV configuration is given by assigning the same non-vanishing VEV  
\equ{
3\, |\langle s^\prime_{3,n} \rangle|^2 = 
6\, |\langle s^\prime_{4,n} \rangle|^2 = 
\sfrac{10}{3}\, |\langle s^\prime_{5,n} \rangle|^2 = 
\frac{e^{2\gvf}}{4\gp^2} \, \frac{\text{Vol}(X)}{\ell_s^8}~,
\nonumber \\
| \langle s^{\prime\prime}_{1,n} \rangle|^2 = 
\frac {3k}{2k-2}\, \frac{a^2}{\ell_s^6}~,~~ 
| \langle s^{\prime\prime}_{2,n} \rangle|^2 = 
\frac {3k}{2k+2}\, \frac{a^2}{\ell_s^6}~, \label{CICY7862_VEVconfig}
} 
to all copies (labeled by $n$) of the scalars in Table~\ref{tb:CICY7862Singlets}. Taking the limit $k \ra \infty$ results in $\text{Vol}(D_i) = 6\, a^2$\,, cf.\ \eqref{DilScalingCICY7862} and \eqref{DivVolCICY7862}. In combination with \eqref{CICY7862_VEVconfig} we infer that the VEVs in the observable sector tend to zero in this limit. 
The second line of \eqref{CICY7862_VEVconfig} shows that the VEVs in the hidden sector are related to the K\"ahler moduli and hence need to be close to the string scale.

%
%
Therefore, the latter VEVs signify a large modification of the original line bundle. Hence, it should be described as a non-Abelian vector bundle. This in turn means that we need to determine the non-Abelian bundle and check its stability, which is difficult. Instead, we only focus on the VEVs in the hidden sector, since the VEVs in the observable sector become negligible in the large $k$ limit. From the weights $w$ of the hidden VEVed states, 
\equa{ \label{VEVed_weights}
\arry{rcl}{
w(s_{1,n}^{\prime\prime}) &=& (0, 0^4, ~1,~0,\sm1)~, 
\\[1ex] 
w(s_{2,n}^{\prime\prime}) &=& (0, 0^4,\sm1,~0,\sm1)~, 
}
}
we see that neither their sum nor difference are roots. Hence, the commutator of the E$_8\times$E$_8$ generators corresponding to these states does not contribute to any non-Abelian F- or D-terms, see \eqref{FandDfrom10D}. This gives a strong (but not sufficient) indication that the vanishing F- and D-terms in the EFT could actually guarantee the relation of the presented VEV-configuration to a stable non-Abelian vector bundle.

In ref.~\cite{Buchbinder:2013dna} the construction of infinite sets of line bundle models on the same geometry (i.e.\ on the tetra-quadric) is investigated. In their case, the authors consider classes of S(U(1)$^5$) bundles in the observable sector where the DUY conditions are fulfilled without charged non-Abelian singlet VEVs. In such a setting, they find that the number of line bundle models should be finite deep inside the K\"ahler cone. However, as they observe, it is difficult to exclude that infinite sets of models are possible when approaching the boundary of the K\"ahler cone.

%
%
\subsection{EFT Breakdown for the infinite set of Schoen models}

For the infinite set of models on the Schoen manifold we encounter similar issues, hence our discussion will be brief. Again, the set of models constructed admits perturbative D- and F-flat directions keeping the K\"ahler moduli well inside the K\"ahler cone. For that, we shift all inherited and exceptional divisor volumes in an appropriate way (as in \eqref{DivVolCICY7862} for the CICY example) to accommodate the one-loop correction in hidden sector, i.e.
\begin{subequations} 
\equ{ 
\text{Vol}(E_r) = \text{Vol}_E + \frac 14\,{N^{\prime\prime}(E_r)}\, \frac{e^{2\gvf}}{8\pi^2}\, \frac{\text{Vol}(X)}{\ell_s^2}~,\\[1ex]
\text{Vol}(\widetilde E_r) = \text{Vol}_E + \frac 14\,{N^{\prime\prime}(\widetilde E_r)}\, \frac{e^{2\gvf}}{8\pi^2}\, \frac{\text{Vol}(X)}{\ell_s^2}~,\\[1ex]
\text{Vol}(R_i) = \text{Vol}_R + \frac 14\,{N^{\prime\prime}(R_i)}\, \frac{e^{2\gvf}}{8\pi^2}\, \frac{\text{Vol}(X)}{\ell_s^2}~,
}
\end{subequations} 
while $\text{Vol}_R >  4 \text{Vol}_E$ is still fulfilled. Then, an admissible D- and F-flat VEV configuration is given by
\begin{subequations} 
\equ{ \label{Schoen_VEVConfig}
|\langle s^\prime_{1,n} \rangle|^2 =  
|\langle s^\prime_{2,n} \rangle|^2 = 
|\langle s^\prime_{3,n} \rangle|^2 = 
\frac{1}{4} \, \frac{e^{2\gvf}}{8\gp^2} \, \frac{\text{Vol}(X)}{\ell_s^8}~,
\\[1ex] 
|\langle s^{\prime\prime}_{4,n} \rangle|^2 =  
|\langle s^{\prime\prime}_{5,n} \rangle|^2 = 
\frac{3}{8}
\frac{1}{4k^2-1}\, \frac{\text{Vol}_R}{\ell_s^6}~.
}
\end{subequations} 
The VEVed singlets are given in Table~\ref{tb:SchoenSinglets} which uses the U(1) charge basis 
\equa{ \label{SchoenU1Basis}
\hspace{-.5ex}
t^\prime_{(1)} &= (\sm1, 1, 0^6)~,~&
t^\prime_{(3)} &= (1, 1, 0^2, 2,0^3)~,~
\\[1ex] \non 
t^\prime_{(2)} &= (0^2, \sm1, 1, 0^{4} )~,~&
t^\prime_{(4)} &= (\sm1,\sm1, 0^2, 1, \sm3, \sm3, 3)~,~
\\[1ex] \non 
t^{\prime\prime}_{(0)} &= (2^3, 0^{5})~,~&
t^{\prime\prime}_{(x)} &= (0^3, 0^{x-1}, 2, 0^{5-x} )~,
}
with $x=1,\ldots,5$. 
This background also satisfies the extended F- and D-flatness conditions~\eqref{FandDfrom10D}, which are presumably a prerequisite for the existence of a stable non-Abelian vector bundle.
The VEVs $\langle s^\prime_{n} \rangle$ in the observable sector are of one-loop order and will vanish for large $k$,  assuming a $k$-dependent dilaton as in \eqref{DilScalingCICY7862} for the CICY example. 
The VEVs in the hidden sector $\langle s^{\prime\prime}_{n} \rangle$ are now suppressed by $1/k^2$. 
However, this does not necessarily imply that this class of infinite models tends back to a line bundle background when $k \rightarrow \infty$, because the number of singlets developing tree-level VEVs grows cubically with $k$, see Table~\ref{tb:SchoenSinglets}.

%
%
\begin{table}
\begin{center}
\scalebox{.9}{
\renewcommand{\arraystretch}{1.1}
\begin{tabular}{|c|c|c||c|c|c|}
\hline 
\multicolumn{3}{|c||}{Observable singlet VEVs} & 
\multicolumn{3}{c|}{Hidden singlet VEVs} 
\\
Mult. & U$^\prime$(1)$^{4}$ charges & Label &
Mult. & U$^{\prime\prime}$(1)$^{6}$ charges & Label 
\\ \hline\hline 
16 & \,1,~0,~3,~0 & $\langle s^\prime_1 \rangle$ & 
$8k(4k^2-1)$ & ~0,~0,~0,-2,~0,-2~ & $\langle s^{\prime\prime}_4 \rangle$ 
\\
16 & \,0,-1,~2,-5  & $\langle s^\prime_2 \rangle$ &
$8k(4k^2-1)$ & ~0,~0,~0,~2,~0,-2~ & $\langle s^{\prime\prime}_5 \rangle$ 
\\
16 &  -1,~0,~3,~0~ & $\langle s^\prime_3 \rangle$ & & & 
\\ \hline 
\end{tabular}
\renewcommand{\arraystretch}{1}
}
\end{center}
\caption{\label{tb:SchoenSinglets} 
This Table lists only the non-Abelian singlets that get VEVed on the Schoen manifold, characterized by their U(1) charges in either the observable or the hidden sector using the U(1) basis~\eqref{SchoenU1Basis}.
}
\end{table}

\section{Discussion}
\label{sc:Conclusion}

%
%
We showed that the fundamental consistency conditions for heterotic CY compactifications, i.e.\ the Bianchi identities and flux quantization, often allow for an infinite set of solutions. It is even possible to have infinite replicas of GUT (and therefore also MSSM-like) models which only differ in the hidden sector. We demonstrated this explicitly for GUT models on the CICY 7862 and on the Schoen manifold. Hence, we believe that this is, in fact, a rather generic feature of line bundle models.

%
%
In the explicit models we saw that we were never able to fully reconcile the growing of an infinite set of models with all known constraints. We could satisfy the DUY equations inside the K\"ahler cone  at the price of having more and more hidden singlets with VEVs that ensure F- and D-flatness. This reflects the general EFT expectation that such infinite sets of models are impossible: Having an arbitrary large number of massless fields in a quantum field theory is problematic even when they reside in a hidden sector, since operators that arise beyond the truncation of the EFT at lowest order in $\alpha^\prime$ become important.

%
%
Our results may seem in conflict with the statement that the number of stable vector bundles with fixed total Chern class $c(\cV)=(\text{rk}(\cV), c_1(\cV), c_2(\cV),c_3(\cV))$ is finite~\cite{Maruyama:1981,Langer:2004}. This number is counted by the Donaldson-Thomas invariants~\cite{Donalson:1996,Thomas:2000} (see the discussion in e.g.~\cite{Anderson:2010ty,Anderson:2014gla}). Nevertheless, one should be careful applying this principle to our setting, since the total Chern class of the hidden sector depends on $k$, as we infer from the $k$-dependent spectrum.

%
%
The effective SUGRA description with line bundle backgrounds should be viewed as an approximation to full string theory compactifications. Therefore, it is instructive to see whether it is possible to construct an infinite set of models in exact string backgrounds. This never seems to be possible: For example, in orbifold constructions one can make the model input data (gauge shift and Wilson lines) arbitrary large. However, this does not lead to an infinite set of different models, rather, the same model is reproduced an infinite number of times.

%
%
Some further insight into whether such infinite sets of line bundle models can be realized in string theory can be obtained from Gauged Linear Sigma Models (GLSMs). The GLSMs corresponding to CICYs in the standard embedding are well established as (2,2) models on the worldsheet. The line bundle backgrounds on non-compact orbifold resolutions can be lifted to (0,2) GLSMs~\cite{Nibbelink:2010wm}. 
Trying to do the same for line bundle backgrounds on compact CYs described as CICYs or toroidal orbifold resolutions never seem to lead to consistent worldsheet theories outside the (2,2) locus: The hypersurface constraints are implemented as worldsheet F-terms associated with Fermi multiplets, which results in anomalous worldsheet gauge theories. These gauge anomalies can only be cancelled if (0,2) chiral multiplets are introduced, whose charges balance those of the Fermi multiplets. But this allows for further worldsheet superpotential interactions, which means that one departs from an Abelian gauge background in target space by constructing more general monad bundles. More importantly, the worldsheet anomaly cancellation conditions only allow for a finite number of solutions in this case. 

%
%
The effect of infinite sets of models is relevant for the interpretation of model statistics. This observation is independent of whether one can construct infinite sets of models in the full string theory or not. On the one hand, one can find more and more very similar models by extending the range of input parameters further and further. Hence, the meaning of model statistics in a finite parameter range is unclear. On the other hand, all known field-theoretic constraints on the line bundle data, which we have discussed here, do not seem to provide any precise conditions on the allowed parameter range of the input data. Consequently, it remains elusive at which point one leaves the regime of validity of the EFT approach.

\begin{acknowledgments}
We would like to thank Lara Anderson, Ralph Blumenhagen, Andrei Constantin, James Gray and Andre Lukas for useful discussions and correspondence. 
The work of F.R.\ was supported by the German Science Foundation (DFG) within the Collaborative Research Center (SFB) 676 ``Particles, Strings and the Early Universe''. The research of P.V. was done and financed in the context of the ERC Advanced Grant project ``FLAVOUR'' (267104) and was also supported by the DFG cluster of excellence ``Origin and Structure of the Universe'' (www.universe-cluster.de). O.L. acknowledges the support by the DAAD Scholarship Programme ``Vollstipendium f\"{u}r Absolventen von deutschen Auslandsschulen'' within the ``PASCH--Initiative''. 
\end{acknowledgments}

\end{document}